\documentstyle[twocolumn]{mn}

\newcommand{\be}{\begin{equation}}
\newcommand{\ee}{\end{equation}}
\newcommand{\ba}{\begin{eqnarray}}
\newcommand{\ea}{\end{eqnarray}}
\newcommand{\mincir}{\raise -2.truept\hbox{\rlap{\hbox{$\sim$}}\raise5.truept 
\hbox{$<$}\ }} 
\newcommand{\magcir}{\raise -2.truept\hbox{\rlap{\hbox{$\sim$}}\raise5.truept 
\hbox{$>$}\ }} 
\newcommand{\minmag}{\raise-2.truept\hbox{\rlap{\hbox{$<$}}\raise 6.truept\hbox 
{$>$}\ }}
\input{psfig.sty}
\title[Zero metallicity stellar sources and the 
reionization epoch]
{Zero metallicity stellar sources and the 
reionization epoch }
\author[P. Cojazzi, A. Bressan, F. Lucchin,
O. Pantano and M. Chavez]{Paolo Cojazzi$^{1}$,
Alessandro Bressan$^{2}$,
Francesco Lucchin$^{1}$,
Ornella Pantano$^{3}$
\newauthor and Miguel Chavez$^{4}$\\
$^{1}$ Dipartimento di Astronomia, Universit\`{a} di Padova, vicolo
dell'Osservatorio 5, I--35122 Padova, Italy\\
$^{2}$ Osservatorio Astronomico, vicolo
dell'Osservatorio 5, I--35122 Padova, Italy\\
$^{3}$ Dipartimento di Fisica {\em Galileo Galilei}, Universit\`{a} di
Padova, via Marzolo 8, I--35131 Padova, Italy\\
$^{4}$ INAOE,  Apartado Postal 51 y 216, Puebla, M\'exico\\}

\begin{document}

\maketitle

\begin{abstract}

We reconsider the problem of the cosmological reionization due to
stellar sources. Using a method similar to that developed by Haiman \&
Loeb (1997), we investigate the effect of changing the stellar models
and the stellar spectra adopted for deriving the ionizing photon
production rate.
In particular, we study the consequences of adopting zero metallicity stars, 
which is the natural choice for the first stellar populations.
We construct young isochrones representative of Population
III stars from existing sets of evolutionary models
(Forieri 1982; Cassisi \& Castellani 1993) and calculate
a suitable library of zero metallicity
model atmospheres.
The number of ionizing photons
emitted by such a zero metal population is about
40\% higher than that produced by standard metal poor
isochrones. 
We find that adopting  suitable zero metallicity models modifies the
reionization epoch. However the latter is still largely affected by
current uncertainties in other important physical processes such as
the efficiency of the star formation and the fraction of escaping UV
photons.

\end{abstract}


\section{Introduction}
 
In the recent years there has been a lot of theoretical work related to the
cosmological reionization.
The source of the ionizing UV background is still uncertain: the
more studied scenarios are those connected to a first generation of stars in
mini--galaxies or to the radiation from massive black holes in small halos.
The Next Generation Space Telescope will be able to observe directly the first 
luminous objects and discriminate between the previous scenarios.
In addition, future satellite experiments (such as MAP and PLANCK) are likely 
to detect the CMB secondary anisotropies due to the reionized
intergalactic medium.

Detailed numerical simulations of the reionization phenomena requires a 
suitable treatment of different physical phenomena: gas dynamics, cooling
processes and radiative transport
(Abel et al. 1998, Norman et al. 1998, Razoumov \& Scott 1999, Gnedin 1999).
However, the complete study of the formation of first objects and the 
feedback effect on the surrounding medium is still out of the possibility
of the present numerical approaches. 
For this reason a number of analytical or semianalytical methods
(e. g. Fukugita \& Kawasaki 1994, Liddle \& Lyth 1995, Tegmark, Silk \&
Blanchard 1994,
Haiman \& Loeb 1997,  hereafter HL97, Valageas \& Silk 1999, Ciardi et al.
1999, Chiu \& Ostriker 1999) have been used for estimating the reionization
epoch and how this depends on the cosmological model and on the formation 
history of ionizing sources.  The effect related to the 
cosmological parameters  are relatively understood in comparison to those 
used for estimating stellar feedback.

In this paper we focus mainly on how the properties of the stars,
in particular the metallicity and stellar winds,
could affect the reionization epoch. In section \ref{sezmod} we briefly 
describe the reionization model. In section \ref{sezpop}
we discuss the peculiarity of the Population III (hereafter PopIII) stars and
present our isochrones and our
spectral library. Finally, in section \ref{sezfin} we present and
discuss our results.

\section{The reionization model}
\label{sezmod}

To study the process of the cosmological reionization we have developed
a code based on the analytical method described in HL97.
The mass function of the dark matter halos is computed according to
the Press--Schechter theory (Press \& Schechter 1974). The  formation of a
stellar population is allowed in any halo with a total mass
$M \ge 10^8 \left[ (1+z) /10 \right]^{-3/2}$M$_\odot$, assuming
that all the H$_2$
is dissociated by the very first radiation sources as soon as
reionization begins.
 
In such  halos  a fraction $f_*$
of the  gas is converted into stars in a single instantaneous burst
({\em simple stellar population} model, hereafter SSP). Given the lack of 
complete  theory of star formation, this 
fraction is assumed to be a universal constant and
it is fixed on the base of the metallicity observed in the 
Lyman--$\alpha$ forest.
If we assume that the carbon observed in these systems is  produced
by the same primordial mini--galaxies we are considering here,
the efficiency of star formation is likely confined in the range 
$0.015 \le f_{\star} \le 0.15$ (see HL97 for more details).
We use the Salpeter IMF, with a low mass limit equal to $0.024\,M_{\odot}$.
Given such a limit, the mass fraction of  stars in the range
$3-8\,M_{\odot}$ is the same as that given by the Scalo IMF, 
used by HL97, so that we 
can refer to their values for $f_{\star}$.

Part of the ionizing photons (with energy $\ge 13.6$ eV) is absorbed by
the gas in the mini--galaxies. This effect is described  by 
introducing a factor which represents the escape fraction $f_{esc}$ of photons.
Due to the incertitude in the estimate of this parameter, we explore the range 
$f_{esc}=0.2 \div 1$.

We assume that the emission is isotropic
and that the intergalactic medium is homogeneous; in this case the volume of the spherical HII
region formed around each galaxy  may be computed 
analytically (see Shapiro \& Giroux 1987).
Integrating over all the sources, we can calculate, at any  redshift,
the filling factor of the HII regions ($F_{HII}$).
The reionization redshift $z_{rei}$ is defined by $F_{HII}(z_{rei})=1$.

\section{Stellar populations of zero metallicity}
\label{sezpop}

In order to compute the SSP photon rate production
$\epsilon_\nu(t,Z)$, we integrate the contribution of stars along an
isochrone of given age, t and metallicity, Z. If $f_\nu
[L(M,t,Z),g(M,t,Z),T_{eff}(M,t,Z)]$ is the spectrum of the star with
initial mass $M$,  luminosity $L$, surface gravity $g$ and effective
temperature $T_{eff}$ along the isochrone of age t and metallicity Z,
and $\psi(M)$ is the initial mass function, then the emissivity is
given by
\begin{equation}
\epsilon_\nu(t,Z)=\int_{M_{inf}}^{M_{sup}} f_\nu [L(M),g(M),T_{eff}(M)]
\psi(M)dM\; .
\label{eps}
\end{equation}
We have neglected, for simplicity, the dependence from age and metallicity 
while the dependence on the gravity has been
explicitly included because, in general, the initial mass is
different from the current stellar mass due to mass-loss along the
isochrone. The procedure is similar to the one adopted in Bressan,
Chiosi \& Fagotto (1994, hereafter BCF94).

Equation \ref{eps} requires the knowledge of the theoretical isochrone
(L(M) , g(M), T$_{eff}$(M) ) and of  suitable model atmospheres.
In the following we discuss our choice for the zero metallicity 
stellar populations.

\subsection{Theoretical isochrones for zero metallicity stars}

The first generation of stars forms from the primordial material,
whose chemical composition is entirely dictated by the cosmological
nucleosynthesis. Detailed computations of primordial
nucleosynthesis set an upper limit for the CNO mass abundance 
$Z_{CNO}\leq10^{-12}$ (Applegate, Hogan \& Scherrer 1987). Only
in the case of inhomogeneous nucleosynthesis and 
baryonic matter density parameter $\Omega_b$=1,
significantly higher values, $Z_{CNO}\leq10^{-9}$, are predicted
(Kawano et al. 1991). These extreme models  
will not be considered here.

The very low abundance of CNO elements affects both the radiative
opacity and the rate of nuclear energy production, as shown in early
studies by Ezer (1961), Ezer \& Cameron (1971), Hartquist \& Cameron
(1977), Bond, Carr \& Arnett (1983), El Eid, Fricke \& Ober (1983) and
Castellani, Chieffi \& Tornamb\'e (1983). 

For very young stellar populations, which are most interesting here,
the largest  effect is due to the lack of the CNO cycle in the most
massive stars because, due to  the very shallow temperature dependence
of the proton--proton reaction rates, very high central temperatures
and  densities are needed in order to halt the gravitational
contraction and set the star onto the ZAMS. Eventually, in stars
more massive than a critical mass ($10\, {\rm M}_\odot \div 20\, {\rm
M}_\odot$, depending on the assumed initial helium abundance), the
central temperature overcomes $10^8$ K, where the triple alpha
reaction becomes effective in producing fresh carbon ($^{12}$C). The
new synthesized $^{12}$C immediately feeds the CNO cycle which begins
to compete with the p--p cycle and becomes the dominant process
when the  CNO abundance reaches a few times $10^{-11}$.
Because of the steep temperature dependence of the CNO reaction rate,
the central temperature and density start decreasing, but the former
never falls below $10^8$ K, so that the CNO abundance keeps
growing under the effect of the triple alpha reaction. At the end of the
central hydrogen burning phase the CNO abundance in massive stars is
of the order of a few times $10^{-9}$, still some orders of magnitude
less than that typical in the most metal poor stars in our Galaxy.
As a consequence the main sequence of zero metallicity
massive stars is definitely hotter (and slightly more luminous) than that
obtained by adopting the lowest abundance found in the metal poor globular
clusters.
\begin{figure}
\centerline{\psfig{file=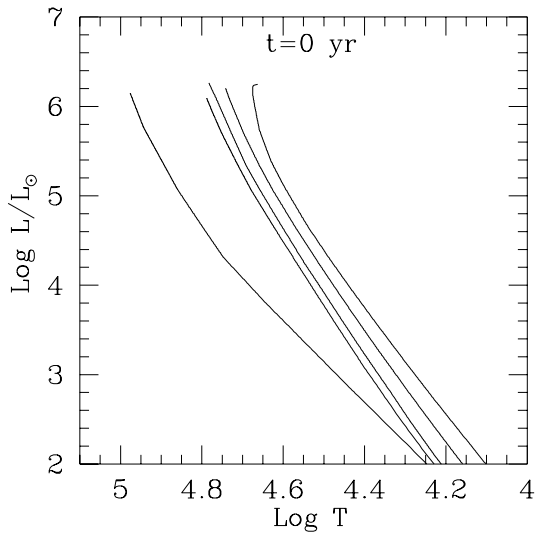,width=7.5cm,height=7.5cm}}
\caption{Zero age main sequence for different initial chemical
composition. Starting from the rightmost one, the
initial composition is Z=0.02, Y=0.28;
Z=0.004,Y=0.24;
Z=0.0004, Y=0.23 (Bertelli et al. 1994) and Z=0.0001, Y=0.23 (Girardi
et al. 1996). The leftmost line is our Z=0 isochrone.}
\label{zcomp}
\end{figure}

\begin{figure}
\centerline{\psfig{file=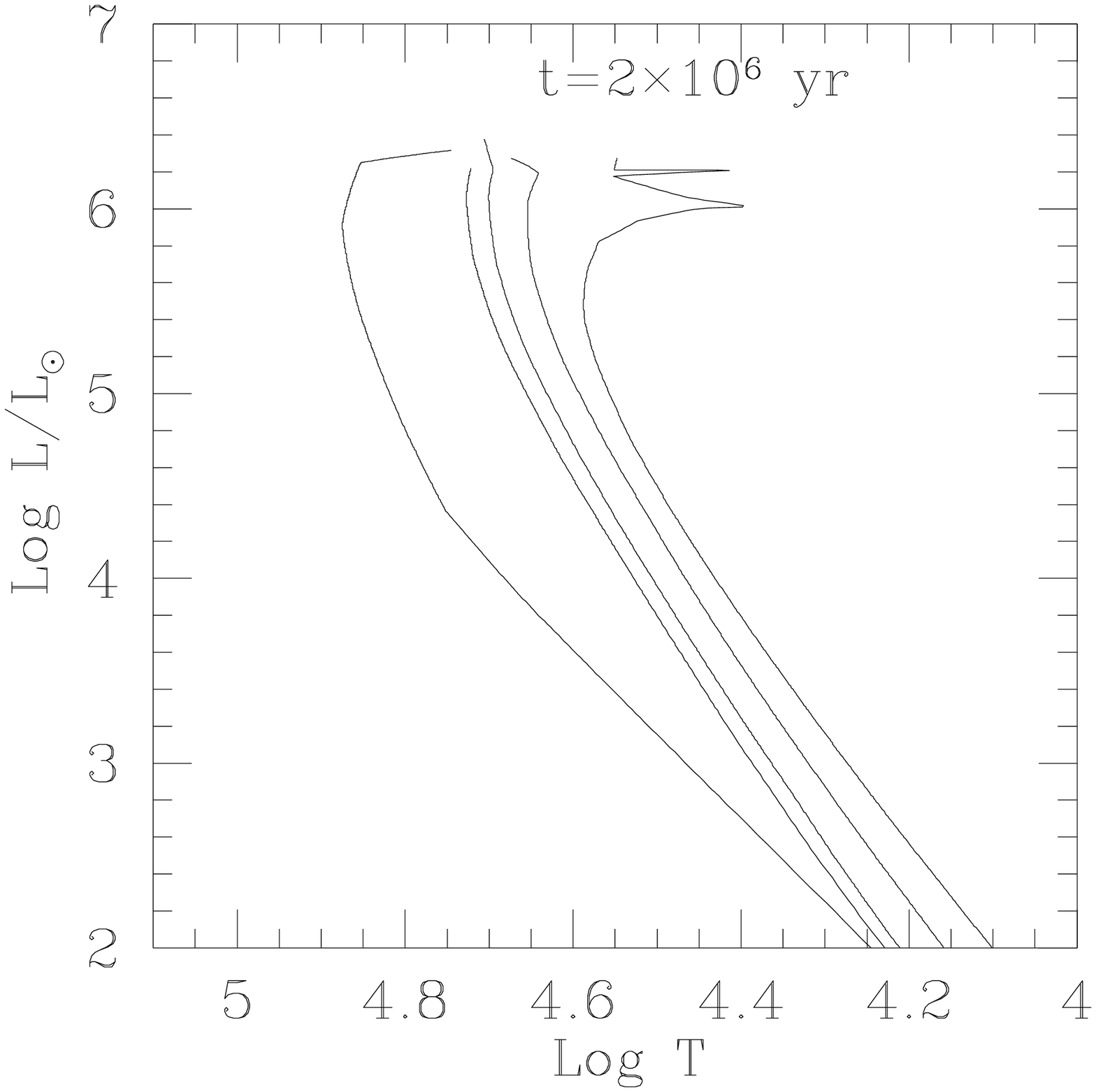,width=7.5cm,height=7.5cm}}
\caption{As in Fig. \ref{zcomp} but for a time t=2Myr.}
\label{zcomp1}
\end{figure}

In this paper zero metallicity isochrones were constructed by combining
sets of stellar evolutionary tracks computed by several authors. 
The first set
is by  Cassisi \& Castellani (1993) and refers to stars between 1.0
M$_\odot$ and 30 M$_\odot$, for a chemical composition
Z=10$^{-10}$. Models of higher mass have been computed by Forieri
(1982) (20 M$_\odot$, 60 M$_\odot$ and 100 M$_\odot$ with Z=0) and by
El Eid et al. (1983) ( M $\geq$ 80 M$_\odot$). Here we use
stellar evolution tracks computed by the former author because of
the finer time resolution of the corresponding tables.
Slightly bluer isochrones are obtained by using the tracks of El Eid et al.
1983 instead of those by Forieri 1982 (Cojazzi et al. 2000).

Isochrones were computed by considering equivalent evolutionary phases
along tracks of different mass, as described in Garcia--Vargas, 
Bressan \& Diaz (1995), in order to carefully follow the
ionizing photon output rate from the most massive stars.
The resulting zero metallicity isochrones are shown in Fig.
\ref{zcomp} and \ref{zcomp1}, for two selected ages, t=0 and t=2 Myr,
respectively. 
In the figures we also show for comparison the 
isochrones of non zero metallicity
(Bertelli et. al 1994, Girardi et. al 1996) in order to highlight
the large effect produced
by the lack of the CNO cycle mainly in the range of massive zero metallicity
stars.
In both figures the metallicity increases from left to right and it is
Z=0, Z=0.0001, Z=0.0004, Z=0.004 and Z=0.02, respectively.

\subsection{Spectral library}

To compute the emissivity of PopIII stellar population
we need a grid of suitable atmospheric models.
Existing models of very low metallicity have been computed by
several authors (see e.g.  Kurucz 1993 and later revisions
and Schmutz et al. 1992). 
The lowest metallicity in the LTE plane parallel models computed by
Kurucz corresponds to [M/H]=-5. The highest temperature in this library
is 50000K and it is significantly lower than that reached by our main
sequence models. On the other hand, Schmutz et al. (1992) computed non LTE
wind models of pure helium stars with temperature ranging from
about 30000K to 150000K. The
latter models were meant to represent the atmospheres of Wolf--Rayet
stars which are characterized by large mass outflows. 
However, if the mass loss rate decreases with the metallicity as
discussed by Kudritzki et al. (1987), it should have a negligible effect in the
case of zero metallicity stars and  high temperature plane parallel models
should be  preferred.

Given the lack of such models, we computed a grid of  fluxes for
zero metallicity atmospheres using the code developed by Auer \&
Heasley (1971).
The parameter space covered by the new grid extends from 10000 to
140000 K in effective temperature and from 4.0 dex to 6.0 dex in surface
gravity. The code is based on the classical assumptions of radiative
and local thermodynamic equilibria, and the plane parallel
approximation. For the computation we have considered 71 optical depth points,
ranging from $-7.0 < \tau_{\rm ROSS} < 2.4$. A total of 45 wavelength
points have been considered extending from 113 {\AA}
to about 4.5
$\mu$. The coverage in
wavelength is sufficient to make the grid suitable for continuum studies since the
Balmer and Lyman breaks at 3646 and 912 {\AA} are among the most
prominent features on the spectral energy distributions in the
parameter space covered by the grid. For a more detailed description
of the grid the reader is referred to Chavez \& Cardona (2000; in preparation).

\begin{figure}
\centerline{\psfig{file=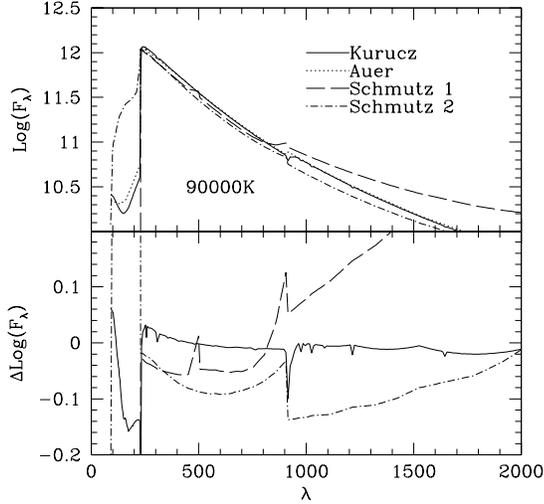,width=7.5cm,height=7.5cm}}
\caption{Model spectra of very metal poor stars at T$_{eff}$=90000 K.
Solid line:  [M/H]=-5  model obtained with  the Kurucz code;
dotted line: zero metallicity model computed with the
Auer \& Heasley (1971) code; dashed line: Schmutz et al. (1992) model
with moderate mass-loss rate; dot--dashed line: Schmutz et al. (1992)
model with strong mass-loss rate. The lower panel depicts the 
logarithmic difference with respect to the Auer \& Heasley model.}
\label{staratmo}
\end{figure}

We also extended the [M/H]= -5 Kurucz grid to higher temperatures, in order
to highlight possible differences between true zero metal atmospheres
and very metal poor ones. 
Fig. \ref{staratmo} compares very hot spectra 
(T$_{eff}$=90000K) in the region near and below the
Lyman break, from the different sets of atmospheric models.
The model computed with the Kurucz code (solid line) and the model
computed with the Auer \& Heasley (1971) code (dotted line) are almost
superimposed. At this temperature the Lyman break is practically
absent. Also the blanketing by metals in the [M/H]=-5 Kurucz grid is
very inefficient. Moreover in both models the flux below about 228 {\AA}
falls rapidly because, in those models,  the abundance of HeII ions
is always high enough to block the corresponding ionizing radiation.

To illustrate this point we plot in the same figure the models
of Schmutz et al. (1992). 
These pure helium non LTE models account for the effect of 
stellar wind and show that enough  HII ionizing 
radiation can escape the star in the case of strong mass-loss (model 1),
the opposite of what occurs in plane parallel models. 
In the case of a less strong wind (model 2), the spectrum is more similar
to the plane parallel case, with a  lack of photons shortward of 228 \AA.

Because very low metallicity stars considered here
likely do not suffer of strong mass-loss rates,  
the use of plane parallel atmospheres
seems thus justified. 
In any case the only significant difference would be in the amount of
HeII ionizing radiation which is able to escape the stellar {\it
photosphere}, which is not relevant in the present context. By
contrast, the possibility that PopIII stars may significantly
ionize HeII (e.g. Tumlinson \& Shull 2000) must be
considered with some caution.

Finally we remark that  we have also considered, for sake of
comparison, two cases with non zero metal content, namely Z=0.0001 and
Z=0.004. In both cases we used the Padova theoretical isochrones, but
Kurucz atmospheres for the former and the {\em CoStar} library
(Schaerer \& De Koter 1997) for the latter.

\section{Discussion and conclusions}
\label{sezfin}

We now consider the effects of zero metallicity stars on the
reionization redshift adopting the model described in section
\ref{sezmod} and the library of stellar isochrones obtained in section
\ref{sezpop}. 

According to our assumptions, we model a galaxy  with an instantaneous
burst of star formation and neglect the local metal enrichment.
Therefore the metallicity is kept constant during the whole process of
reionization. 
The effect of metal enrichment due to merging with already evolved
systems where the gas has been  polluted  by previous stellar
generations, it is not considered here and, in this respect, the results
obtained with zero metallicity stars must be considered as upper
limits to the reionization redshift.

\begin{figure}
\centerline{\psfig{file=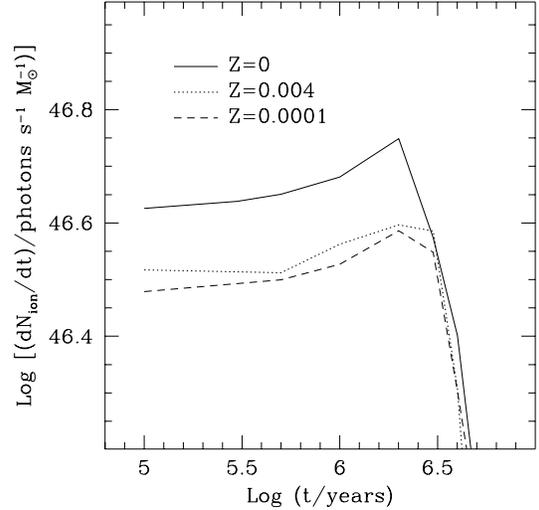,width=7.5cm,height=7.5cm}}
\caption{Ionizing photon production rate as a function of the 
age for simple stellar populations of three different metallicities:
Z=0 (solid line), Z=0.0001 (dotted line) and Z=0.004 (dashed line).}
\label{phot}
\end{figure}

Fig. \ref{phot} compares the evolution of 
the ionizing photon production rate of an instantaneous burst of star formation
at different metallicities. It can be seen from the figure that
the number of ionizing photons
emitted by young zero metallicity SSPs is always from 30\% to 40\% higher than
that obtained with typical low metallicity populations.

As far as  the assumptions underlying the cosmological model are concerned,
we considered two typical cases:  a standard CDM model with parameters
and a flat CDM model with non zero cosmological constant.
The former is taken as reference model while the latter
is suggested by observations (Viana \& Liddle 1996).
The characteristic parameters of the models are:
($\Omega_0$,
$\Omega_{\Lambda}$, $\Omega_b$, $h$, $\sigma_{8 h^{-1} {\rm Mpc}}$,
$n$, $\Gamma$) = (1, 0, 0.05, 0.5, 0.52, 1, 0.45) for the CDM and
($\Omega_0$, $\Omega_{\Lambda}$, $\Omega_b$, $h$,
$\sigma_{8 h^{-1} {\rm Mpc}}$, $n$, $\Gamma$) = (0.3, 0.7, 0.03, 0.83,
0.93, 1, 0.21) for the $\Lambda$CDM.

In both models the power spectrum of primordial fluctuations has been
normalized to fit the present cluster abundance (Eke,
Cole \& Frenk 1996).
For each cosmological model, we investigated two extreme values of the
gas conversion fraction, $ f_*=0.015$ and  $f_*=0.15$, and
of the photon escape fraction, $f_{esc}=0.2$ and $f_{esc}=1$.
These values can be considered as representative of the uncertainty
with which the corresponding processes are known.

\begin{table}
\centering
\caption[] {Reionization redshifts $z_{rei}$ in
($\Omega_0$, $\Omega_{\Lambda}$, $\Omega_b$, $h$, $\sigma_{8 h^{-1} {\rm Mpc}}$,
$n$, $\Gamma$) = (1, 0, 0.05, 0.5, 0.52, 1, 0.45)
a CDM model, for
different values for $Z$, $f_*$ and $f_{esc}$.}
\tabcolsep 4pt
\begin{tabular}{clccccc} \hline \hline              
&f$_{*}$ &    \multispan{2}{\hfill \quad 0.015 \hfill}  & \multispan{2}{ \hfill 0.150
\hfill}\\
& f$_{esc}$  &\quad 0.2 &  1 \quad & \quad 0.2 & 1\\
Z&&&&&\\
\hline  
0.004  \hfill  \vline  & & \quad   4.9  &  8.6 \quad & \quad    9.9  &    12.7 \\
0.0001 \hfill  \vline  & & \quad   5.2  &  8.8 \quad & \quad   10.1  &    12.8 \\
 0     \hfill  \vline  & & \quad   6.2  &  9.5 \quad & \quad   10.8  &    13.4 \\
\hline
\label{scdm}
\end{tabular}
\end{table}
\begin{table}
\centering
\caption[]{As in Table \ref{scdm} in a $\Lambda$CDM model
($\Omega_0$, $\Omega_{\Lambda}$, $\Omega_b$, $h$,
$\sigma_{8 h^{-1} {\rm Mpc}}$, $n$, $\Gamma$) = (0.3, 0.7, 0.03, 0.83, 0.93, 1,
0.21).}
\tabcolsep 4pt
\begin{tabular}{clccccc} \hline \hline              
&f$_{*}$ &    \multispan{2}{\hfill \quad 0.015 \hfill}  & \multispan{2}{ \hfill 0.150
\hfill}\\
& f$_{esc}$  &\quad 0.2 &  1 \quad & \quad 0.2 & 1\\
Z&&&&&\\
\hline  
0.004  \hfill  \vline  & & \quad   7.1   &  11.9 \quad & \quad   13.6  &   17.2 \\
0.0001 \hfill  \vline  & & \quad   7.5   &  12.1 \quad & \quad   13.8  &   17.4 \\
 0     \hfill  \vline  & & \quad   8.7   &  13.0 \quad & \quad   14.7  &   18.1 \\
\hline
\label{lcdm}
\end{tabular}
\end{table}

The results are summarized in Table \ref{scdm} for the CDM model 
and in Table \ref{lcdm} for the $\Lambda$CDM model.
Because of the enhanced ionizing photons production rate, the
reionization is faster when PopIII stars are included, leaving  all
the others parameters unchanged.
In particular, comparing the case of PopIII stars with that of Z=0.004 stars
we see that the relative variation in the reionization redshift
$z_{rei}$ is about 20\% in the models with a late reionization and
reduces to about 5\% in the case of an early reionization.
In some cases the late reionization obtained with low metallicity SSPs
is not compatible with the lower limit on $z_{rei}$ inferred from
observations of recent limits on the Gunn-Peterson effect at redshift
$\leq$5 (e.g. Songaila et al. 1999), and observations of high redshift
Lyman--$\alpha$ emitters (Hu, Cowie \& McMahon 1998) that would not be
detected in presence of a neutral intergalactic medium
(Miralda-Escud\'e 1998). We notice that,
in such cases, the adoption of PopIII stars alone  is sufficient
to reconcile the models with the observations.

Turning now to the favorite cosmological model ($\Lambda$CDM) we see
that the redshift of reionization is confined in the range between $9 \leq z
\leq 18$. The main uncertainties are  introduced  by the parameters
$f_*$ and $f_{esc}$, which roughly imply a variation of the photon
production rate by a factor of 10 and 5, respectively.
From our analysis it appears that,
for a given cosmological model, these are the factors which actually define
the reionization epoch from stellar sources.

\section* {Acknowledgments}  

We warmly thank Santi Cassisi for kindly providing unpublished and new
PopIII evolutionary models and Daniel Schaerer for kindly providing
the {\em Costar} spectral library. Lauro Moscardini is acknowledged
for useful discussions. We warmly thank the referee, A. Weiss, for his
comments that greatly helped us to improve upon the first version of
the paper. This work was partially supported by Italian MURST. A.B.
and P.C. acknowledge the support by the European Community under TMR
grant ERBFMRX--CT96--0086.

\end{document}